\documentclass[aps,pre,twocolumn,groupedaddress,showpacs,floatfix]{revtex4}
\usepackage{amsmath}
\usepackage{amssymb}
\usepackage{graphicx}
\usepackage{epic}
\usepackage{eepic}
\usepackage{subfigure}
\begin{document}

\title{Comparative analysis of protein structure using multiscale
additive functionals}
\author{Marconi Soares Barbosa$^1$, Rinaldo Wander Montalv\~ao$^2$,
Tom Blundell$^2$ and Luciano da Fontoura Costa$^1$}
\affiliation{$^1$Institute of Physics at S\~ao Carlos,
University of S\~ ao Paulo, S\~{a}o Carlos,
SP, PO Box 369, 13560-970, 
Phone +55 16 3373 9858, 
FAX +55 162 71 3616, 
Brazil, 
marconi@if.sc.usp.br}
\affiliation{
$^2$Department of Biochemistry,
University of Cambridge, 
80 Tennis Court Road,  Cambridge, CB2 1GA,
Phone +44 1223 333 628,
FAX +44 1223 766 082
United Kingdom
}

\date{\today}

\begin{abstract} 
This work reports a new methodology aimed at describing
characteristics of protein structural shapes, and suggests a framework
in which to resolve or classify automatically such structures into
known families. This new approach to protein structure
characterization is based on elements of integral geometry using
biologically relevant measurements of shape and considering them on a
multi-scale representation which align the proposed methodology to the
recently reported \emph{tube picture} of a protein structure as a
minimal representation model. The method has been applied with good
results to a subset of protein structures known to be especially
challenging to revert into families, confirming the potential of the
proposed method for accurate structure classification.
\end{abstract}

\pacs{}

\maketitle

\section{Introduction}
Evolution has produced a huge number of protein families and
super-families whose members possess similar sequences and
three-dimensional structures. Restraints on evolutionary divergence
are mainly related to the protein function, and therefore selective
pressure tends to operate on the three-dimensional
structure~\cite{Bajaj:1984}. The HOMSTRAD~\cite{HOMSTRAD} is an
example of a database of protein structures organized into homologous
families.  As a consequence of the global proteomic effort, the number
of known structures is growing at an impressive rate and has passed
the total of 39000 structures. It is remarkable progress but, on the
other hand, it also introduces an overwhelming amount of data to be
manually classified on those databases. With more than 400 structures
solved every month, the challenge for automatic protein structural
comparison and classification is greater than ever. Most of the
protein comparison methods depend mainly upon structural alignment and
RMSD measures, and therefore are not completely
reliable~\cite{Koehl:2001}. While RMSD is a good measure of structure
similarity for almost identical proteins, it cannot be used to judge
dissimilarity since it violates the triangle inequality. It means that
any system based on RMSD alone is unable to cluster structures and,
consequently incapable of classifying them into families. In addition,
the reliance on sequence alignments introduces a drawback because it
is virtually impossible to avoid errors during the alignment
construction.

In this paper we investigate the potential of an algorithm adapted to
automatically classify proteins into HOMSTRAD families. This algorithm
is based on concepts of Integral Geometry~\cite{Stoyan:1995}, know as
Morphological Image Analysis (MIA), which has been recently applied to
a series of problems due to its simplicity in design and
implementation. Fields as diverse as Neuroscience~\cite{Barbosa:2003b}
and Materials Sciences~\cite{Raedt01} have benefited from this
approach.

\section{Additive Shape Functionals}\label{AF} 

We start by describing the mathematical aspects of the adopted
procedure.  The Minkowski functionals of a body $K$ in the plane are
proportional to the familiar geometric quantities of area $A(K)$,
perimeter $U(K)$ and the connectivity or Euler number $\chi(K)$.  The
usual definition of the connectivity from algebraic topology in two
dimensions is the difference between the number of connected $n_c $
components and the number of holes $n_h$, $\chi(K)=n_c-n_h$. In the
Euclidean space, there are two kinds of holes to consider.  First, we
have the pure hole, a completely closed region of white voxels
surrounded by black voxels. Second, the tunnels.  The Euler
characteristic is consequently given as $\chi(K)= n_c -n_t +n_h$,
where $n_t$ is the number of tunnels and $n_h$ is the number of pure
holes.  There is an additional geometric quantity to consider in the
three-dimensional space, namely the mean curvature or breadth
$B(K)$. By exploring the additivity of the Minkowski functionals,
their determination reduces to counting the multiplicity of basic
building blocks that disjointly compose the object. For example a
voxel can be decomposed as a disjointed set of 8 vertices, 12 edges, 6
faces and one open cube. The same process can be applied to any object
in a lattice. For a three-dimensional space, which is our interest
regarding protein structures, see~\cite{Raedt01,Barbosa:2003b}, we
have
\begin{align}
&V(\mathcal{P})=n_3, \quad S(\mathcal{P})=-6n_3+2n_2, \\ \nonumber 
&2B(\mathcal{P})=3n_3-2n_2+n_1,\quad \chi(\mathcal{P})=-n_3 +n_2 -n_1+n_0,
\end{align}
Where $n_3$ is the number of interior cubes, $n_2$ is the number of
open faces, $n_1$ is the number of sides and $n_0$ is the number of
vertices. So, the procedure to calculate Minkowski functionals of a
pattern $\mathcal{P}$ can be reduced to counting the number of
elementary bodies of each type that compose a voxel (cubes, faces,
edges and vertexes) belonging to $\mathcal{P}$.

\section{Protein structure, tube picture and multiscale signatures}

The protein structure in our approach is defined essentially by the
geometrical/topological nature of its backbone. All $\alpha$-carbon
atom coordinates are identified from a \emph{.pdb} file and an
interpolation scheme is used to connect neighboring atoms by a
straight path. This design procedure attaches a variable resolution to
the method, as the highly refined atomic scale data has to be
truncated during the process.

In our analysis the calculation of the Minkowski functionals are
incorporated into a multiscale framework. In such a scheme, all four
quantities are calculated as a function of a control parameter as some
transformation is made on the structure of interest. In this paper we
consider this transformation to be the process of exact dilations and
the control parameter the dilation radius. Our choice is particularly
suited as the exact dilation procedure naturally fits itself in what
has been described as the \emph{tube picture} for protein structure
analysis~\cite{Banavar:2003}, a minimalist biophysical reasoning of
the protein model.  While the intricate aspects of the
geometry/topology are accounted for at each spatial scale by the
Minkowski functionals, the space surrounding the backbone is probed by
performing the dilation of the structure and this information is
condensed in what we call henceforth multiscale signatures.  The
behavior of such signatures, particularly the topologically related
ones, can be discontinuous. For example the process of dilation may
change abruptly the number of pure holes or tunnels at particular
scales and these facts are registered for all scales in the multiscale
signature for the connectivity or Euler number (characteristic).

\section{Results and Discussion}
\setlength{\subfigtopskip}{0pt}
\setlength{\subfigcapskip}{0pt}
\setlength{\subfigbottomskip}{0pt}
\begin{figure}[htbp]
\begin{center}
\setlength{\tabcolsep}{-3pt}
\renewcommand{\baselinestretch}{0.5}
\begin{tabular}{c}
\includegraphics*[scale=0.26,angle=-90]{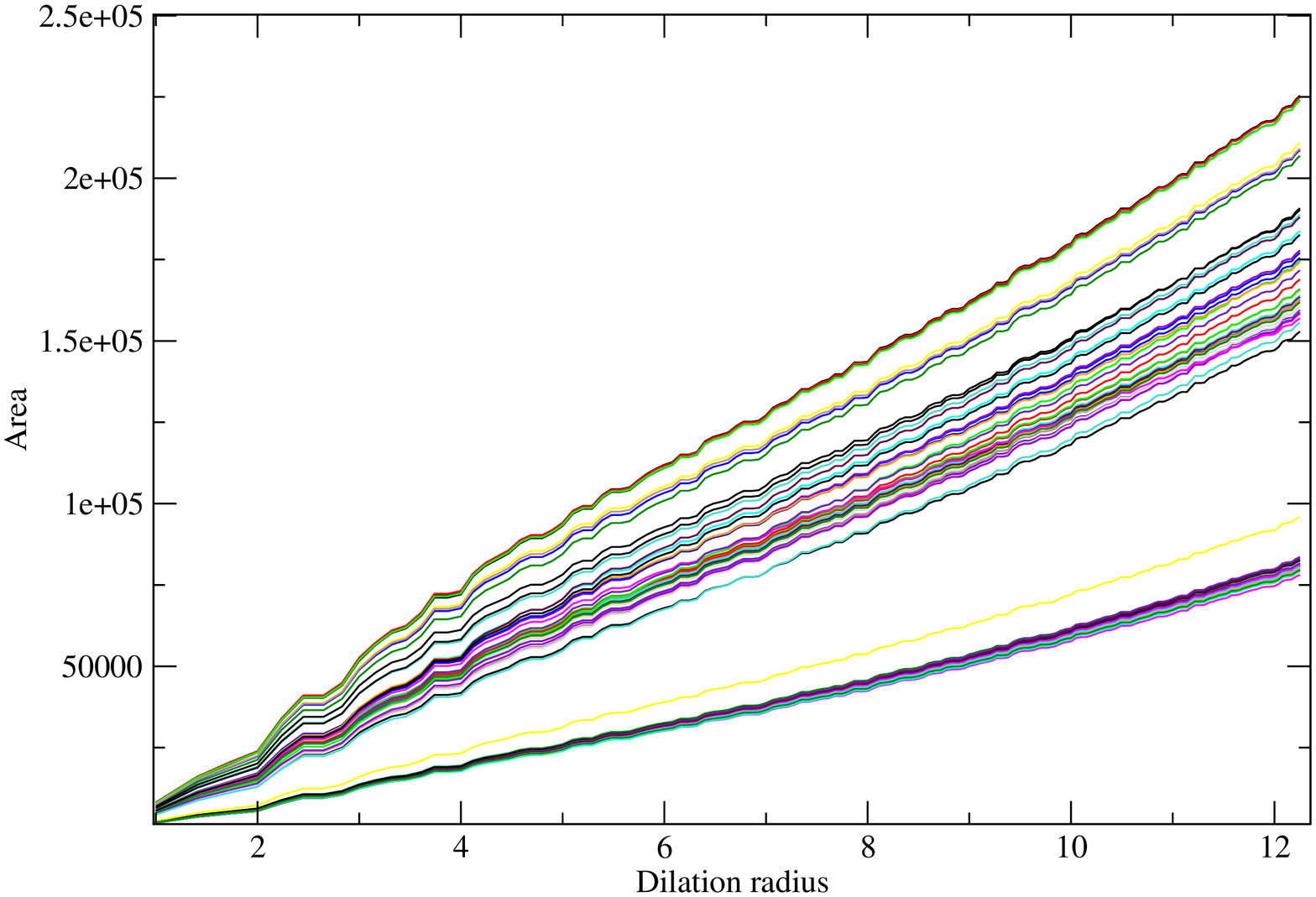}\\
\includegraphics*[scale=0.26,angle=-90]{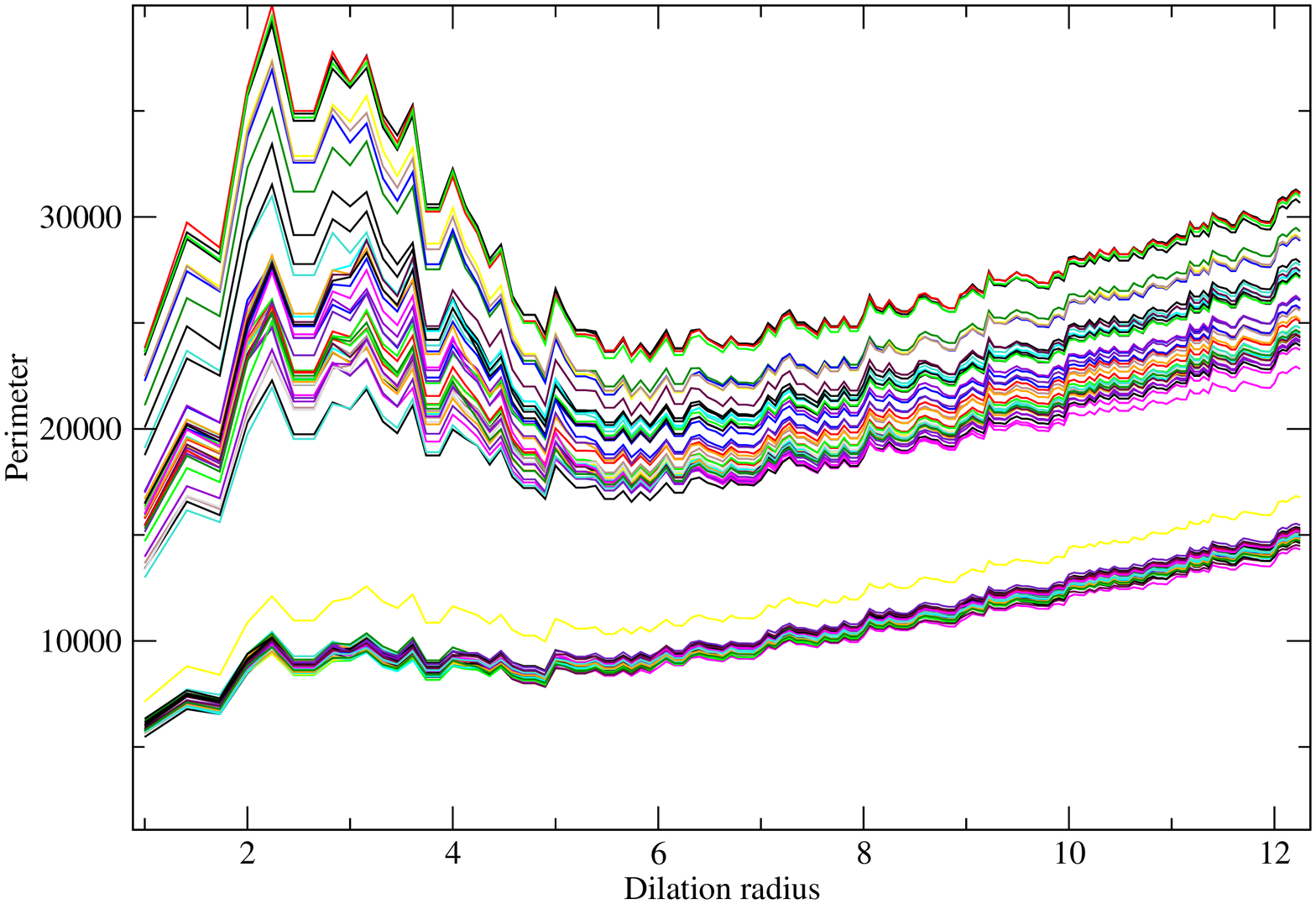}\\
\includegraphics*[scale=0.26,angle=-90]{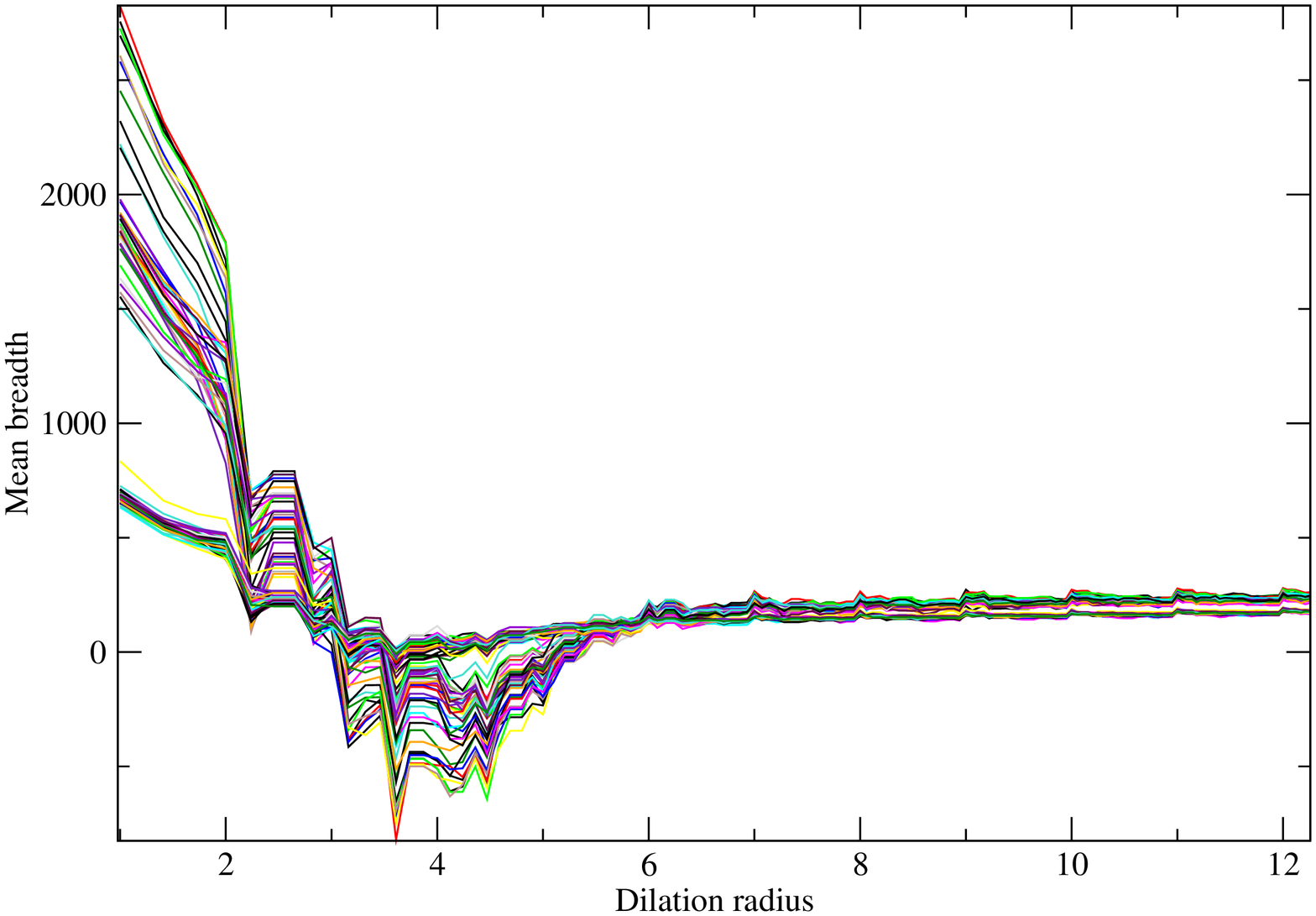}\\
\includegraphics*[scale=0.26,angle=-90]{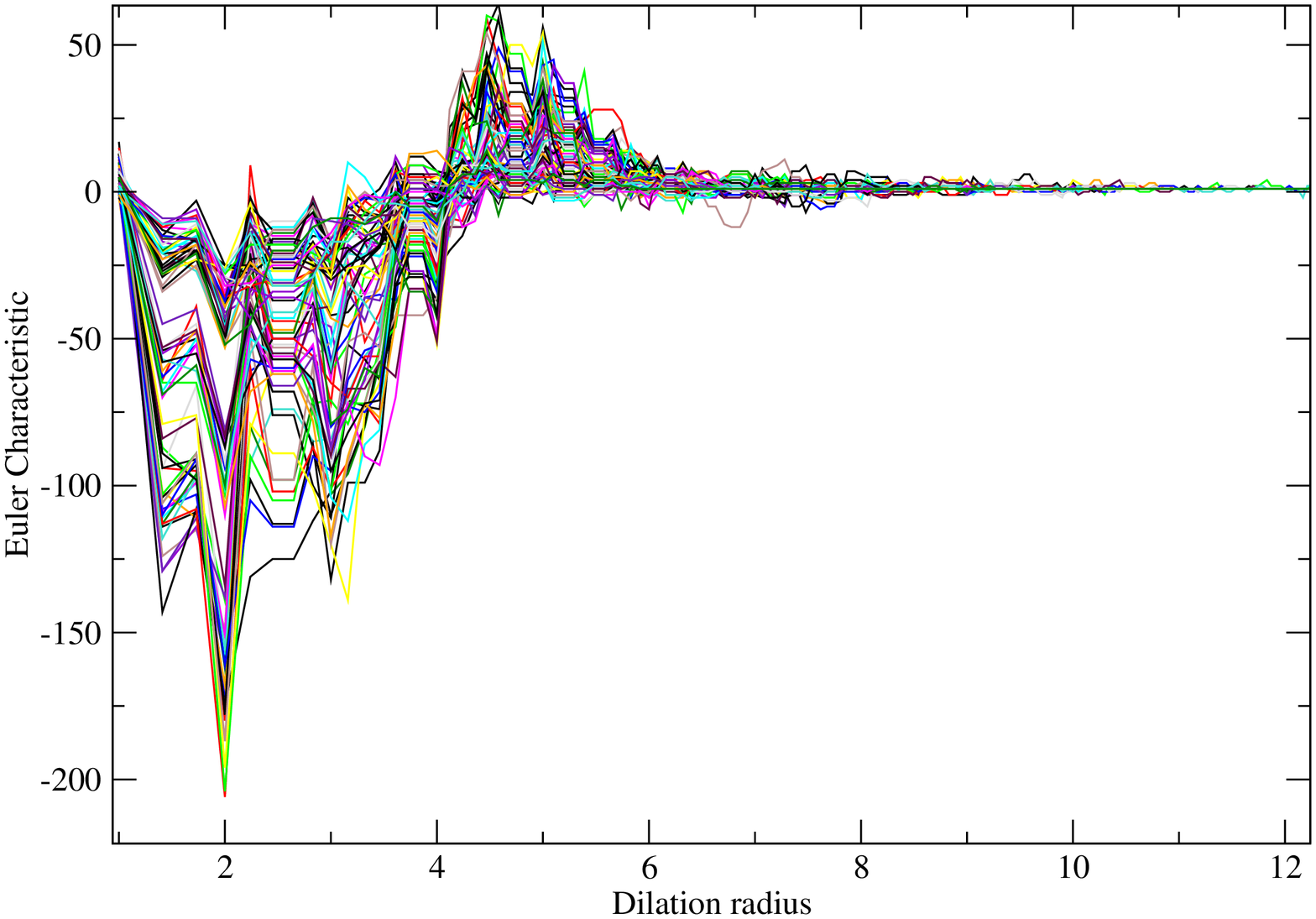} 
\end{tabular}
\caption{\label{fig:signatures} Multiscale signatures associated with the four
  Minkowski functionals in the Euclidean space.}
\end{center}
\end{figure}

Figure~\ref{fig:signatures} shows all considered functionals
signatures for a set of 71 proteins which were chosen specifically
because of their similarities. The range of scales shown in these
graphs encompasses the initial structure and the final filled volume
without holes and tunnels ($\chi=1$). There are both similarities and
striking differences whose subtleties, until now, have been handled
only by more complex algorithms.

For each of those signatures in Figure~\ref{fig:signatures} we select
three features in an attempt to globally characterize the structure
and, by doing so, minimize the amount of data needed for future
classification based on Minkowski functionals. For the signatures of
Area and Perimeter, we evaluated the standard deviation, its integral,
and the scale at which the integral of the curve reaches half of the
total value. For the signatures of the Connectivity and of the Mean
Breadth we measured the standard deviation, the integral of the curve
and the monotonicity index given by $i=(i_s+i_d+i_p)/i_s$ where
$i_{s,d,p}$ are the counts for each time the curve increase, decrease
or stay constant.
\begin{figure}[htb]
 \begin{centering} \includegraphics*[scale=1.4,angle=-90]{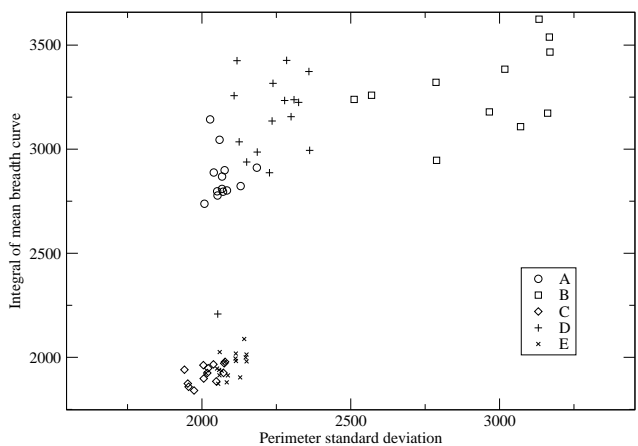}
 \caption{A scatter plot derived from the mean breadth and the perimeter alone leads
   to a discriminative feature space.\label{fig:scatter}}
 \end{centering}
\end{figure}
Table~\ref{tab:cross} shows the numeric results obtained by classical
discriminant analysis~\cite{McLachlan:1992} based on the twelve above
global measures and quantifies the classification potential of the
proposed framework. Such a discriminant analyis projects the
measurements in such a way as to optimize their separation, expressed
in terms of high interclass and low intraclass dispresions. It is
remarkable that, although the structures were specially chosen to make
a reduction into families difficult, this approach managed to
perfectly classify four out of the five families.  A mistake was made
in class C, were it misclassified 1 out of 13 structures. It is
worthwhile to note that although exhibiting different foldings, alpha
plus beta in the class C and all alpha in the class E, their average
length and topological properties in general are quite
similar. Figure~\ref{fig:scatter} shows a two-dimensional section of
the complete feature space defined by measures from the mean breadth
and connectivity only. It provides a more economical discriminating
clustering, albeit with overlaps.
\begin{table}[tbp]
\begin{center}
\begin{tabular}{l|l|l|l|l|c|c|c|}\hline\hline
            & A & B & C & D & E & Error& Posterior.Error \\\hline A
(asp) & 13 & 0 & 0 & 0 & 0 & 0.0000000& 0.0000477 \\\hline B (ghf13) &
0 & 11 & 0 & 0 & 0 & 0.0000000& 0.0000025 \\\hline C (ghf22) & 0 & 0 &
12 & 0 & 1 & 0.0769231& 0.1051906 \\\hline D (kinase) & 0 & 0 & 0 & 16
& 0 & 0.0000000& 0.0247150 \\\hline E (phoslip) & 0 & 0 & 0 & 0 & 18 &
0.0000000& -0.0104105 \\\hline

Overall     &    &     & & & &              0.0140845&      0.0221997 \\\hline\hline
\end{tabular}
\end{center}
\caption{The result of a classical discriminant analysis for the
12 features extracted from the multiscale signatures. \label{tab:cross} }
\end{table}

\section{Conclusion}

In this paper we have accessed the potential of the multi-scale
Minkowski functionals for protein morphological characterization and
structural analysis. We found that these functionals are potentially
suited to this kind of analysis, as substantiated by the results
obtained for a distinct set of structures known to have highly similar
topological features. For all but one family of structures, namely the
glycosyl hydrolase family 22, the classification through a classical
discriminant analysis yielded fully accurate results. These results
are comparable with the best approach so far~\cite{Rogen:2003}, which
uses considerably more parameters and is based on a complex
concept. In addition to the classification result, it is important to
emphasize the simplicity of the algorithm and the clear relationship
between the quantities used for the characterization and familiar
geometrical, topological and biological concepts. This direct relation
to familiar measurements, combined with the simplicity for
implementing the MIA approach, suggests that this kind of analysis is
a particularly useful tool for classifying the shape of protein
structures.
\bibliographystyle{unsrt} \bibliography{protein}
\end{document}